\documentclass[aps,showpacs,twocolumn]{revtex4}

\usepackage{epsfig}
\usepackage{latexsym}
\usepackage{amssymb}

\newcounter{saveeqn}

\begin{document}


\title{Numerical studies of $\Phi^2$-Oscillatons}


\author{
Miguel~Alcubierre$^{1}$, Ricardo Becerril$^{2}$, F.
Siddhartha~Guzm\'{a}n$^{3}$, Tonatiuh~Matos$^{4}$,
Dar\'{\i}o~N\'{u}\~{n}ez$^{1}$, and L.
Arturo~Ure\~{n}a-L\'{o}pez$^{5}$\footnote{Present address: Instituto
de F\'{\i}sica de la Universidad de Guanajuato, A. P. E-143,
C.P. 37150, Le\'{o}n, Guanajuato, M\'{e}xico.}
}

\affiliation{
$^{1}$Instituto de Ciencias Nucleares, Universidad
Nacional Aut\'{o}noma de M\'{e}xico, A.P. 70-543, 04510 M\'{e}xico,
D. F., M\'{e}xico.\\
$^{2}$Instituto de F\'{\i}sica y Matem\'{a}ticas, Universidad
Michoacana, Edif. C-3, Ciudad Universitaria 58040, Morelia,
Michoac\'{a}n, M\'{e}xico. \\
$^{3}$Max-Planck-Institut f\"{u}r Gravitationsphysik, Am
M\"{u}hlenberg 1, D-14476 Golm, Germany. \\
$^{4}$Departamento de F\'{\i}sica, Centro de Investigaci\'{o}n y
de Estudios Avanzados del
IPN, A.P. 14-740, 07000 M\'{e}xico D.F., M\'{e}xico.\\
$^{5}$Astronomy Centre, University of Sussex, Brighton BN1 9QJ,
United Kingdom. 
}


\date{January, 2003}


\begin{abstract}
We present an exhaustive analysis of the numerical evolution of the
Einstein-Klein-Gordon equations for the case of a real scalar field
endowed with a quadratic self-interaction potential. The
self-gravitating equilibrium configurations are called {\it
oscillatons} and are close relatives of boson stars, their complex
counterparts. Unlike boson stars, for which the oscillations of the
two components of the complex scalar field are such that the spacetime
geometry remains static, oscillatons give rise to a geometry that is
time-dependent and oscillatory in nature. However, they can still be
classified into stable (S-branch) and unstable (U-branch) cases. We
have found that S-oscillatons are indeed stable configurations under
small perturbations and typically migrate to other S-profiles when
perturbed strongly. On the other hand, U-oscillatons are intrinsically
unstable: they migrate to the S-branch if their mass is decreased and
collapse to black holes if their mass is increased even by a small
amount. The S-oscillatons can also be made to collapse to black holes
if enough mass is added to them, but such collapse can be efficiently
prevented by the gravitational cooling mechanism in the case of
diluted oscillatons.
\end{abstract}




\maketitle


\section{Introduction}
\label{one}

\vspace{5mm}

In a seminal paper Seidel \& Suen~\cite{seidel91a} found that there
exist non-singular, time-dependent equilibrium configurations of
self-gravitating real scalar fields. These oscillating soliton stars
are called {\it oscillatons}, and are solutions of the
Einstein-Klein-Gordon (EKG) system of equations for minimally coupled
real scalar fields.  The time-dependence of these solutions appears as
a fundamental ingredient that allows singularities to be avoided, in
contrast to static solutions of the EKG equations with real scalar
fields, where singularities frequently appear. The case of real scalar
fields is quite different to the case of complex scalar fields, for
which the EKG equations provide the so-called boson stars, which are
non-singular solitonic solution with a static geometry (the components
of the complex field oscillate in precisely such a way that the
stress-energy tensor is time independent).  However, the stability of
oscillatons has not been studied in as much detail as in the case of
boson stars~\cite{seidel90a,seidel94a,seidel98a}. Such studies are
necessary because of the possible role of oscillatons in astrophysics
and cosmology, where real scalar fields have been proposed as
candidates for the dark matter in the
Universe~\cite{matos01a,alcubierre02a,mielke02a}.

In this paper we want to complement previous
studies~\cite{seidel91a,seidel94a,urena02a,urena02x,hawley02a} on
oscillatons with a numerical analysis of the evolution of the EKG
equations, much in the same way as it has been done for boson
stars~\cite{seidel90a,seidel94a,seidel98a}. The main aim is to
investigate whether oscillatons are stable. For simplicity, we only
consider here the spherically symmetric case of a real scalar field
$\Phi$ endowed with a quadratic scalar potential of the form
\mbox{$V(\Phi)=(m^2/2) \, \Phi^2$}. Other cases will be treated in
future publications.

This paper is organized as follows. In section~\ref{two}, we present
the necessary mathematical background to find equilibrium
configurations and to further evolve the EKG equations. As we shall
see, the equilibrium configurations may be classified in two different
groups: the S and U-branches. The numerical methods and tests are
presented in section~\ref{three}. In sections~\ref{four} and
\ref{five}, we analyze the results obtained from the evolution of the
EKG equations for different oscillatons. We separately study the
behavior of S and U-oscillatons by adding to them small and strong
perturbations that change their total mass. We conclude in
section~\ref{conclusion}.


\section{Mathematical background}
\label{two}

To begin with, we consider the spherically symmetric line element
\begin{equation}
ds^2 = -\alpha^2 dt^2 + a^2 dr^2
 + r^2 \left( d\theta^2 + \sin^2 (\theta) d\varphi^2 \right)
\label{metrica}
\end{equation}

\noindent with $\alpha(r,t)$ the lapse function and $a(r,t)$ the
radial metric function.  We choose the polar-areal slicing
condition (i.e. we force the line element to have the above form
at all times, so that the area of a sphere with $r=R$ is always
equal to $4 \pi R^2$); this choice of gauge will force the lapse
function $\alpha(r,t)$ to satisfy an ordinary differential
equation in $r$. Through out the work, we will be using units such
that $c=\hbar=1$, and we express the gravitational constant in
terms of he Planck mass: $G=1/m_{\rm{Pl}}^2$.

The energy momentum tensor of a scalar field $\Phi$ endowed with a
quadratic self-interaction potential \mbox{$V(\Phi)=(m^2/2) \Phi^2$}
is
\begin{equation}
T_{\mu \nu} = \Phi_{,\mu }\Phi_{,\nu} - \frac{g_{\mu \nu}}{2} \,
\left[ \Phi^{,\alpha }\Phi_{,\alpha} + m^2 \Phi^2 \right] \, .
\label{tensor}
\end{equation}

\noindent The non-vanishing components of $T_{\mu \nu}$ are
\begin{eqnarray}
-{T^0}_0 &=& \rho_\Phi = \frac{1}{2} \left[ \alpha^{-2} {\dot{\Phi}}^2
+ a^{-2} \Phi^{\prime 2} + m^2 \Phi^2 \right] \; ,\label{tensor1} \\
T_{01} &=& {\mathcal P}_{\Phi} = \dot{\Phi} \Phi^\prime \; , \label{tensor2} \\
{T^1}_1 &=& p_r = \frac{1}{2} \left[ \alpha^{-2} {\dot{\Phi}}^2
+ a^{-2} \Phi^{\prime 2} - m^2 \Phi^2 \right] \; , \label{tensor3} \\
{T^2}_2 &=& p_\bot = \frac{1}{2} \left[ \alpha^{-2} {\dot{\Phi}}^2
- a^{-2} \Phi^{\prime 2} - m^2 \Phi^2 \right] \; , \label{tensor4}
\end{eqnarray}

\noindent and also ${T^3}_3 = {T^2}_2$. These different components are
identified as the energy density $\rho_\Phi$, the momentum density
${\mathcal P}_\Phi$, the radial pressure $p_r$, and the angular
pressure $p_\bot$. The parameter $m$ is interpreted as the mass of the
scalar particles. Over-dots denote $\partial/\partial t$ and primes
denote $\partial/\partial r$.

The evolution of the metric functions $\alpha$ and $a$ can be obtained
from the Einstein equations $G_{\mu \nu}=\kappa_0 T_{\mu \nu}$, with
$\kappa_0 = 8\pi G$. In order to write appropriate evolution
equations, we now introduce the first order variables $\Psi
=\Phi_{,r}$ and \mbox{$\Pi=a \Phi_{,t} / \alpha$}. Also, we define the
dimensionless quantities $r=x/m, \, t \rightarrow t/m, \, \Phi
\rightarrow \Phi/\sqrt{\kappa_0}$, where we note that the bosonic mass
is the natural scale for time and distance.

Using these new variables, the Hamiltonian constraint becomes
\begin{equation}
\frac{a_{,x}}{a} = \frac{1-a^{2}}{2x}+\frac{x}{4} \left[ \Psi^{2}
+ \Pi^{2} + a^{2} \Phi^2 \right] \, ,
\label{ham}
\end{equation}
and the polar-areal slicing condition takes the form
\begin{eqnarray}
\frac{\alpha _{,x}}{\alpha } &=&\frac{a_{,x}}{a} + \frac{a^{2}-1}{x}
- x a^{2} \Phi^2  \, .
\label{slice}
\end{eqnarray}

The evolution of the scalar field $\Phi$ is governed by the
Klein-Gordon (KG) equation, which appears as a consequence of the
conservation equations of the scalar field energy-momentum
tensor~(\ref{tensor}) in the form
\begin{equation}
{T^{\mu \nu}}_{;\nu} = \Phi^{,\mu} \left( \Box - m^2 \right)
\Phi = 0,
\end{equation}
where $\Box=g^{\alpha \beta} \nabla_\alpha \nabla_\beta$ is the
d'Alambertian operator. The KG equation is equivalent to the following
set of first order differential equations
\begin{eqnarray}
\Phi_{,t} &=& \frac{\alpha}{a} \, \Pi \, ,
\label{KG1} \\
\Pi_{,t} &=& \frac{1}{x^2}\left( \frac{x^2 \alpha \Psi}{a}\right)
_{,x} - a\alpha \, \Phi \, ,
\label{KG2} \\
\Psi_{,t} &=& \left( \frac{\alpha \Pi}{a} \right)_{,x} \, .
\label{KG3}
\end{eqnarray}

Equations (\ref{ham}-\ref{KG3}) form the complete set of differential
equations to be solved numerically. The evolution equation for $\Pi$
above is further transformed into the equivalent form:
\begin{equation}
\Pi_{,t} = 3 \frac{d}{dx^3} \left( \frac{x^2 \alpha \Psi}{a} \right)
- a\alpha \, \Phi  \, .
\label{KG3-2}
\end{equation}
Notice that the first term on the right hand side of this equation
includes now a first derivative with respect to $x^3$ (and not a third
derivative).  The reason for doing this transformation has to do with
the numerical regularization near the origin of the $1/x^2$ factor in
equation~(\ref{KG2}) above (see Ref.~\cite{hawley00a}).



\subsection{Eigenvalue problem for equilibrium configurations}
\label{two:i}

In order to find the equilibrium configurations of oscillatons,
eqs.~(\ref{ham}-\ref{KG3}) are solved using Fourier expansions for
both the metric and the scalar field
functions~\cite{seidel91a,urena02a,urena02x,hawley02a}. We briefly
describe here the procedure used in~\cite{urena02x} to find such
equilibrium configurations.

In order to deal with the non-linearities present in the EKG
equations, it is convenient to introduce the new variables
$A(t,r)=a^2(t,r), \, C(t,r)=(a/\alpha )^2$, for which
eqs.~(\ref{ham}-\ref{KG3}) take the form
\begin{eqnarray}
A^\prime &=& \frac{A x}{2} \left( C \dot{\Phi}^2 + \Phi^{\prime 2}
+ A \Phi^2 \right) + \frac{A}{x}(1-A) \, , \label{a2} \\
C^\prime &=& \frac{2C}{x} \left[1+A\left(\frac{1}{2} x^2 \Phi^2 -1 \right)
\right] \, , \label{a3} \\
C \ddot{\Phi} &=& - \frac{1}{2}\dot{C} \dot{\Phi} + \Phi^{\prime \prime}
+ \Phi^\prime \left(\frac{2}{x}- \frac{C^{\prime}}{2 C} \right)
- A \Phi \, , \label{akg} \\
\dot{A} &=& x A \dot{\Phi} \Phi^\prime \, . \label{a1}
\end{eqnarray}
The lapse function is later obtained as $\alpha^2
(t,x)=A(t,x)/C(t,x)$.  Notice that Eq.~(\ref{a1}) is a consequence of
the momentum constraint (the $\{t,r\}$ part of the Einstein
equations.)

We shall consider the Fourier expansions
\begin{eqnarray}
\Phi (t,x) &=& \sum^{j_{max}}_{j=1} \phi_j (x)
\cos (j \omega t) \; , \nonumber \\
A (t,x) &=& \sum^{j_{max}}_{j=0} A_j (x) \cos (j \omega t) \; ,
\label{fexp} \\
C (t,x) &=& \sum^{j_{max}}_{j=0} C_j (x) \cos (j \omega t) \; ,
\nonumber
\end{eqnarray}
where $\omega$ is called the fundamental frequency and $j_{max}$ is
the mode at which the Fourier series are truncated. Solutions are
obtained by introducing the Fourier expansions~(\ref{fexp}) in
eqs.~(\ref{a2}-\ref{a1}), and setting each Fourier coefficient to
zero; that is, the EKG equations are reduced to a set of coupled
ordinary differential equations.

The boundary conditions are determined by requiring non-singular and
asymptotically flat solutions, for which equations~(\ref{a2}-\ref{a1})
become an eigenvalue problem. Thus, it is only necessary to determine
the initial values $\phi_i(0), \, C_i(0)$ (the fundamental frequency
is an {\it output} value) corresponding to a given central central
value $\phi_1(0)$, to obtain different equilibrium configurations.

\begin{figure}[htp]
\centerline{ \epsfysize=6cm \epsfbox{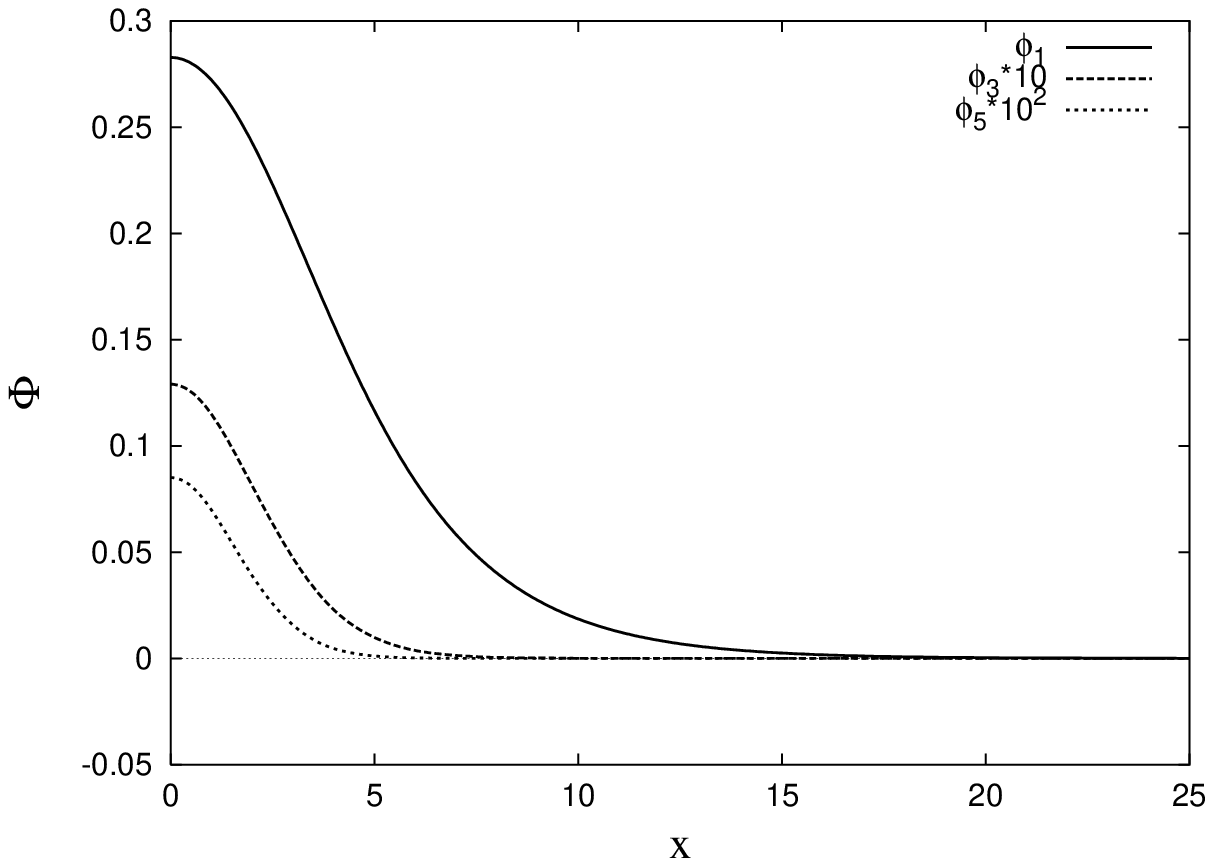}} \centerline{
\epsfysize=6cm \epsfbox{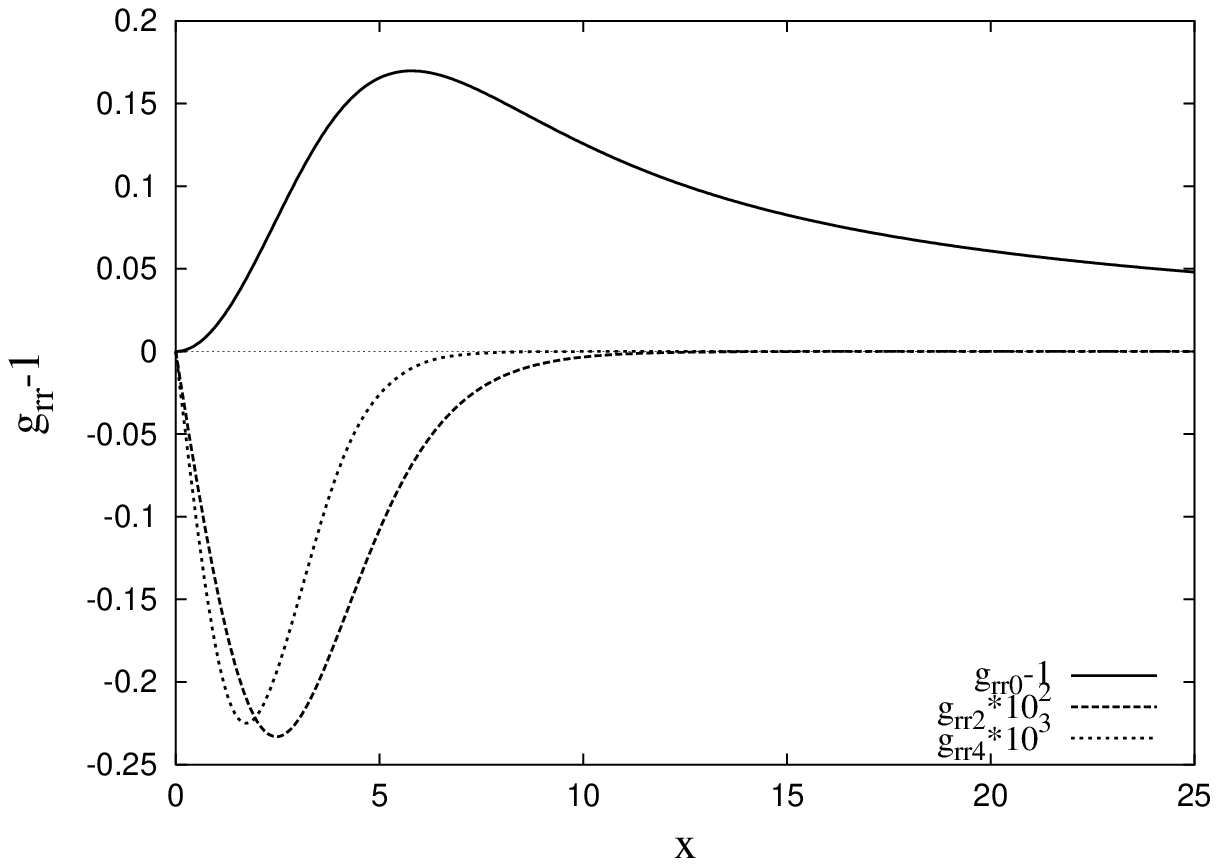}} \caption{Non-zero Fourier
coefficients of the scalar field $\Phi$ and the radial metric
coefficient $g_{rr}-1$ (see eq.~(\ref{fexp})) for a configuration
with $\phi_1(0)=0.2828$. The total mass is $M=0.5726 \, m^2_{\rm Pl}/m$
and the fundamental frequency is $(\omega /m)=0.9128$. The
solution shown here was calculated up to the 6th Fourier mode
($j_{max}=6$ in eqs.~(\ref{fexp})).  The convergence of the
Fourier series is manifest (notice the re-scaling of the higher
Fourier modes).} \label{fig:phi_s}
\end{figure}

A typical oscillaton solution is shown in Fig.~\ref{fig:phi_s}. Even
though we are solving non-linear equations, the Fourier series
converges rapidly. A particular feature of the solutions is that they
are represented only by odd Fourier coefficients of the scalar
field $\Phi$ and the even coefficients of the metric functions $A$ and
$C$.

\begin{figure}[htp]
\centerline{ \epsfysize=6cm \epsfbox{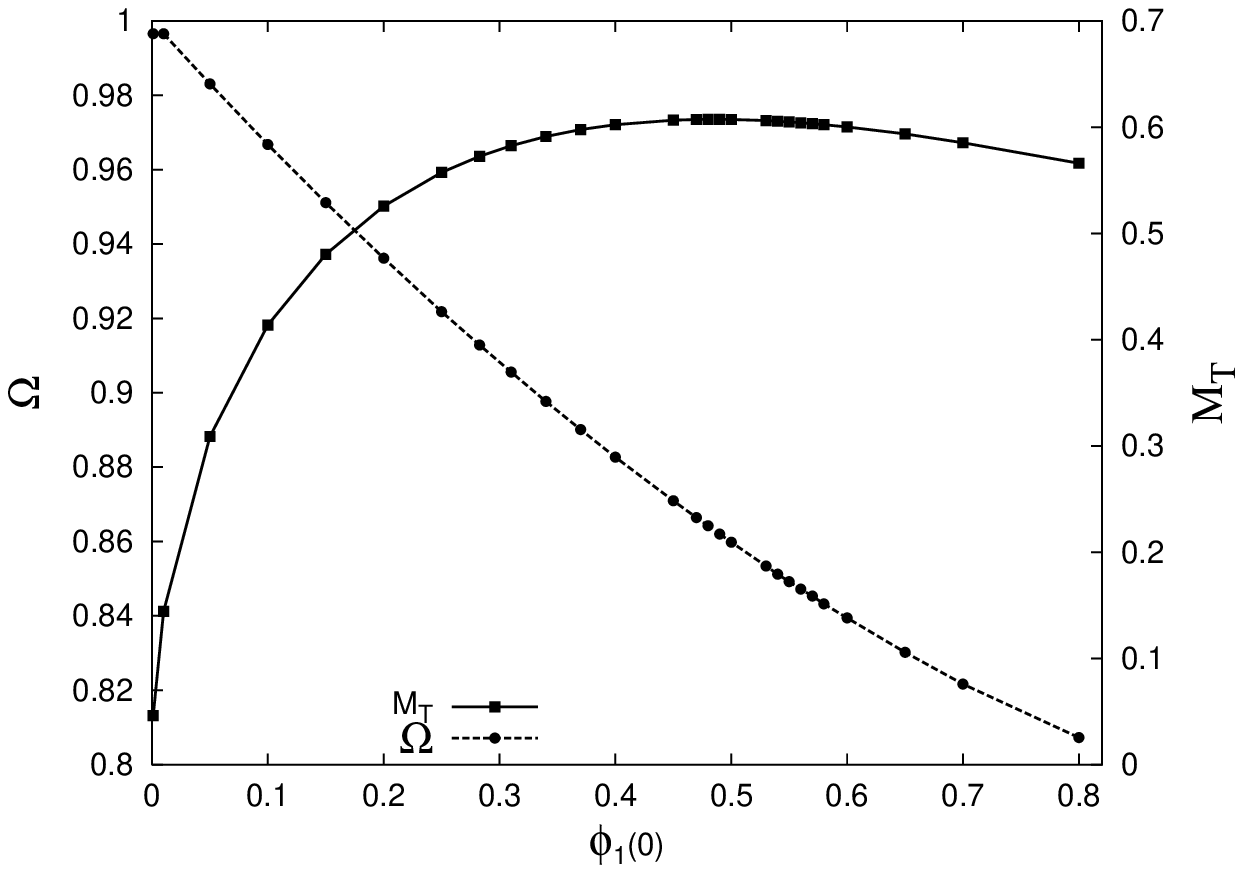}}
\centerline{ \epsfysize=6cm \epsfbox{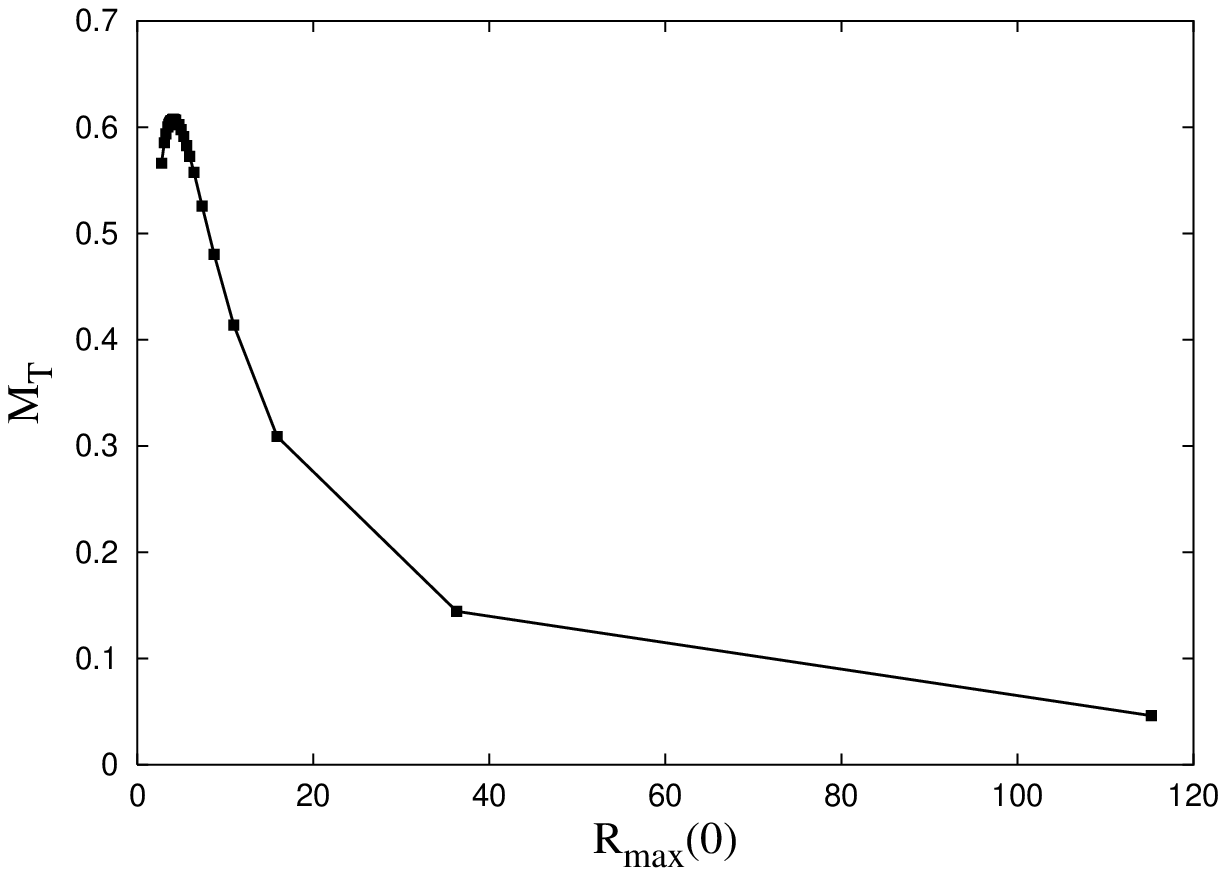}}
\caption{(Top) Total masses $M_T$ (in units of $m^2_{\rm Pl}/m_\Phi$)
and fundamental frequencies ($\Omega$) of different oscillatons. The
critical (maximum) mass is $M_c=0.607 \, m^2_{\rm Pl}/m_\Phi$ for a
configuration with a central value ${\phi_1}_c (0)=0.48$; its
corresponding frequency is $(\omega /m) =0.864$. (Bottom) Plot of the
total mass $M_T \, {\rm vs} \, R_{max}(0)$ (the latter in units of
$m^{-1}$), the radius at which the metric coefficient $g_{rr}$ reaches
its maximum value at $t=0$.}
\label{fig:mass_s}
\end{figure}

In Fig.~\ref{fig:mass_s}, we show the calculated total mass ($M_T$),
the fundamental frequency ($\Omega \equiv \omega /m$) and the radius
at which the radial metric coefficient reaches its maximum value at
$t=0$ ($R_{max}(0)$) for different configurations. In the case of
oscillatons, the position of the maximum of $a^2$ is not a fixed value
but instead oscillates in time. However, as we shall see below (see
Fig.~\ref{fig:mvsr_ec}), the amplitude of such oscillations is quite
small and the initial value can be taken as representative of each
oscillaton.  The maximum mass $M_c = 0.607 \, m^2_{\rm Pl}/m_\Phi$ is
reached for a central value ${\phi_1}_c (0)=0.48$, to which also
corresponds a fundamental frequency $\Omega =0.864$. The fundamental
frequency is always such that $\Omega \leq 1$ for all oscillatons,
with $\Omega =1$ for the trivial solution. In general, more massive
oscillatons oscillate with a smaller fundamental frequency.

\section{Numerics}
\label{three}

\subsection{Numerical algorithm for the evolution of the system}

In order to integrate the Klein-Gordon
equations~(\ref{KG1},\ref{KG2},\ref{KG3}) we used a method of lines
with second order centered differences in space. For the time
integration we use a method inspired in the Iterative Crank Nicholson
(ICN) scheme with three iterations (see for
example~\cite{alcubierre00a,teukolsky00a}).  The standard ICN is used
to integrate a system of evolution equations of the form:
\begin{equation}
\frac{\partial u_i}{\partial t} = S_i (u_j, \partial_{x_k} u_j) \; .
\end{equation}
Given values of the variables $u_i$ at time step $t=n \Delta t$, one
updates the values of the variables to time step $t = (n+1) \Delta t$
in the following way:

\begin{eqnarray}
u_i^{(1)} &=& u^n_i + (\Delta t / 2) \; S^n_i \; , \\
u_i^{(k)} &=& u^n_i + (\Delta t / 2) \; S^{(k-1)}_i \quad k=2,...,N-1 \; , \\
u_i^{n+1} &=& u^n_i + \Delta t  \; S^{(N-1)}_i \; ,
\end{eqnarray}
where $u_i^n = u_i (t=n \Delta t)$, $u_i^{n+1} = u_i (t= (n+1) \Delta t)$,
and $u_i^{(k)}$ are intermediate values. Taking $N=3$ is enough to
obtain a second order accurate, stable
scheme~\cite{alcubierre00a,teukolsky00a} (in fact, taking $N=2$ is
enough for second order accuracy, but it is unstable).

For our purposes we have modified the above algorithm for the case
$N=3$ in the following way:
\begin{eqnarray}
u_i^{(1)} &=& u^n_i + (\Delta t / 3) \; S^n_i \; , \\
u_i^{(2)} &=& u^n_i + (\Delta t / 2) \; S^{(1)}_i \; , \\
u_i^{n+1} &=& u^n_i + \Delta t  \; S^{(2)}_i \; ,
\end{eqnarray}
The reason for this modification is that the above scheme is
considerably less dissipative than standard three-step ICN.  In fact,
for linear equations one can show that the above scheme is third order
in time. We have found our modified scheme ICN to be very stable and
robust in practice.

Once we have advanced the variables $\Phi$, $\Psi$ and $\Pi$ one time
step using the above algorithm, we substitute their new values into
equations~(\ref{ham}) and~(\ref{slice}).  These are simple ordinary
differential equations on the radial coordinate that we solve using a
standard second order Runge-Kutta scheme.


\subsection{Boundary conditions}

Our set of equations is singular at $x=0$.  To avoid the
singularity, we stagger the origin and take a spatial grid of the
form \mbox{$x_i = (i - 1/2) \Delta x$}.  The fictitious point at
$x_0 = -\Delta x/2$ is used to impose appropriate parity
conditions: $\Pi$ is even and $\Psi$ is odd.  Notice that we can
integrate $\Phi$ all the way to the boundary point $x_0 = -\Delta
x/2$, since its evolution equation does not require the evaluation
of spatial derivatives.

At the outer boundary we also need to impose boundary conditions.
Notice again that we do not need to apply a boundary condition for
$\Phi$, as its evolution equation can be integrated all the way to the
boundary point. For $\Pi$ we assume that, for large enough $x$, it
behaves as an outgoing wave pulse of the form:
\begin{equation}
\Pi = u(x-t)/x \; ,
\label{eq:outgoing}
\end{equation}
with $u$ an arbitrary function.  In differential form this becomes
\begin{equation}
\partial_x \Pi + \partial_t \Pi + \Pi/x = 0 \; ,
\end{equation}
which, performing finite difference can be solved for the unknown
boundary value at the new time level.  On the other hand, it is
not difficult to convince oneself that the function $\Psi$ does
not behave as an outgoing wave at the boundary.  However, we are
assuming that $\Pi$ does and, as a consequence, so does $\Phi$. In
particular, the outgoing wave boundary condition applied to $\Phi$
can be seen to imply that at the boundary:
\begin{equation}
\Psi = - \Pi - \Phi/x \; .
\end{equation}
This equation can then be used to obtain boundary values for $\Psi$
once those of $\Phi$ and $\Pi$ are known.

Finally, we need to mention the boundary conditions used for the
ordinary differential equations that have to be solved to find the
metric functions $\alpha$ and $a$.  For $a$, we use the fact that
local flatness implies that $a(x=0)=1$ and $\partial_x a (x=0) = 0$.
These two conditions imply that $a(x_0)=a(x_1)=1 + {\cal O}(\Delta
x)^3$.  We use these two boundary values at the first two grid points
and integrate the second order hamiltonian constraint outwards.

For the lapse function $\alpha$ we use the fact that, on a vacuum, our
slicing condition implies that we are in Schwarzschild coordinates, so
we must have $\alpha=1/a$.  We then assume that our boundaries are
sufficiently far away as to be always in a vacuum, and impose
$\alpha=1/a$ as an outer boundary condition.  The slicing condition is
then integrated inwards.  One could presumably improve on this by
setting $\alpha(x=0)={\rm constant}$, $\partial_x \alpha (x=0) = 0$ as
boundary conditions on the origin, integrating the slicing condition
outwards, and then re-scaling it so that the lapse goes as $1 + k/x$
far away ($k$ is a constant; notice that the slicing condition is
scale invariant).


\subsection{Code tests}

\begin{figure}[htp]
\centerline{ \epsfysize=6cm \epsfbox{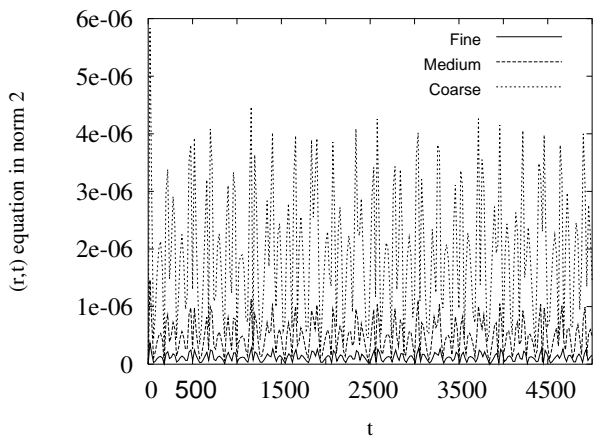}}
\caption{The momentum constraint~(\ref{adot}) for three different
resolutions: Fine ($dr=0.01$), Medium ($dr=0.02$) and Coarse
($dr=0.04$). The initial conditions corresponds to a Gaussian pulse of
the form $\Phi(0,x) = 0.296 \, e^{-x^2/5.35}$, and total mass
$M=0.575$.}
\label{fig:convergence}
\end{figure}

In order to illustrate how all these ingredients work together
properly, we study now the accuracy of our numerical methods in some
particular cases. In Figure~\ref{fig:convergence} we show the
convergence of a system initially consisting of a Gaussian pulse of
scalar field. The equation we use to calibrate our techniques is the
$(r,t)$ component of the Einstein's equations (the momentum
constraint):
\begin{equation}
\beta := a_{,t} - \frac{1}{2} x \alpha \Phi \Pi = 0 \; ,
\label{adot}
\end{equation}
which should be satisfied for an exact solution\footnote{In a previous
work~\cite{seidel91a}, the component momentum constraint was used
to evolve $a$ and the Hamiltonian constraint was monitored to
calibrate the accuracy of the numerical methods. In this
manuscript we have decided to do the opposite}. What we show in
the plot is the $L2$ norm of the value of $\beta$ across the grid as a
function of time for three different resolutions. The initial
Gaussian corresponds to an oscillaton with $M = 0.575$, so that it
is in the stable branch and has a long time life; the boundary is
located at $x=50$ so that the evolutions were carried out up to
100 crossing times. The fact that the value of $\beta$ goes down by a
factor of 4 every time the resolution is doubled shows that the
code remains second order convergent.

\section{Evolution of oscillatons: The S-branch and the quasi-normal modes}
\label{four}

We shall call S-branch oscillatons those equilibrium configurations to
the {\em left} hand side of the critical configuration in the plot $M
\, {\rm vs} \, \phi_1(0)$ in Fig.~\ref{fig:mass_s} (or equivalently,
those configurations to the {\em right} hand side in the plot of $M \,
{\rm vs} \, R_{max}$). As a first step in the study of the stability
of oscillatons, we will numerically evolve these equilibrium
configurations.

We should keep in mind that, compared to realistic oscillatons, the
solutions presented in Fig.~\ref{fig:mass_s} are already
perturbed. These perturbations arise because of the inherent
discretization error of the numerical solutions and (mainly) of the
truncated Fourier expansions. Therefore, the eigen-solutions of
section~\ref{two:i} will be treated as already {\it slightly
perturbed} profiles\cite{seidel91a}.

The main results presented in this section can be summarized as
follows.  i) S-oscillatons are stable against {\it small}
perturbations. ii) Besides the fundamental oscillations of the system,
there is an overall vibration of the oscillaton when slightly
perturbed. The frequencies ($f$) of these vibrations should be
identified with the so-called quasi-normal modes of the system, and the
resulting plot $f \, {\rm vs} \, M$ is a particular feature of the
$\Phi^2$-oscillatons. As we shall see below, such quasi-normal modes
are very important in the study of evolved profiles.

We take the scalar field profiles obtained by solving
eqs.~(\ref{a1}-\ref{akg}) to be the initial data for the evolution
equations.  In this form, the initial conditions for the scalar field
are (see eqs.~(\ref{fexp}))
\begin{eqnarray}
\Phi (t=0,x) &=& \sum^{j_{max}}_{j=1} \phi_j (x) \, , \\
\Phi^\prime (t=0,x) &=& \sum^{j_{max}}_{j=1} \phi^\prime_j (x) \, \\
\dot{\Phi} (t=0,x) &=& 0 \, .
\end{eqnarray}
and then the metric functions $\alpha,a$ are calculated through the
Einstein equations. In this manner, we will also check the consistency
between the eigenvalue problem and the numerical evolution since no
additional information (like the metric functions, the fundamental
frequency $\omega$, etc.) is taken initially.

\begin{figure}[tp]
\centerline{ \epsfysize=6cm \epsfbox{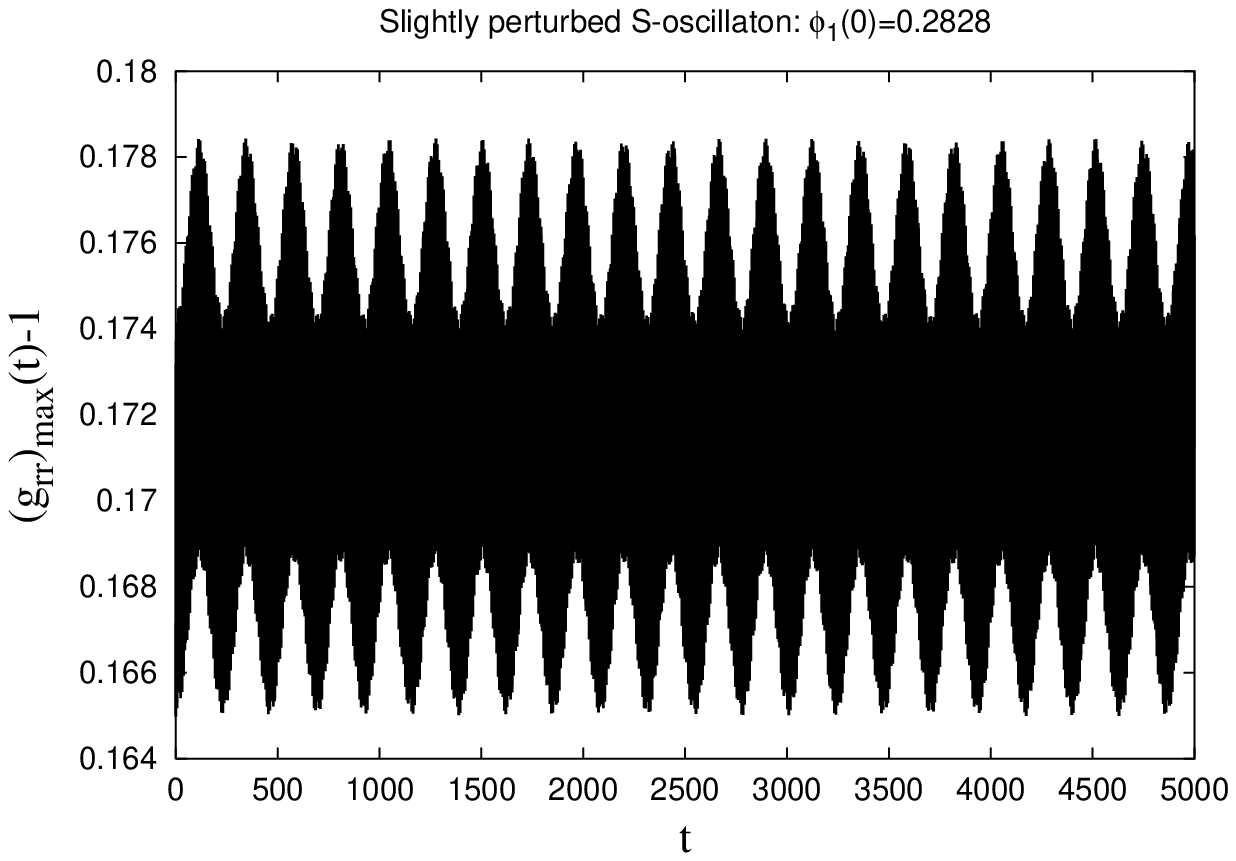}}
\centerline{ \epsfysize=6cm \epsfbox{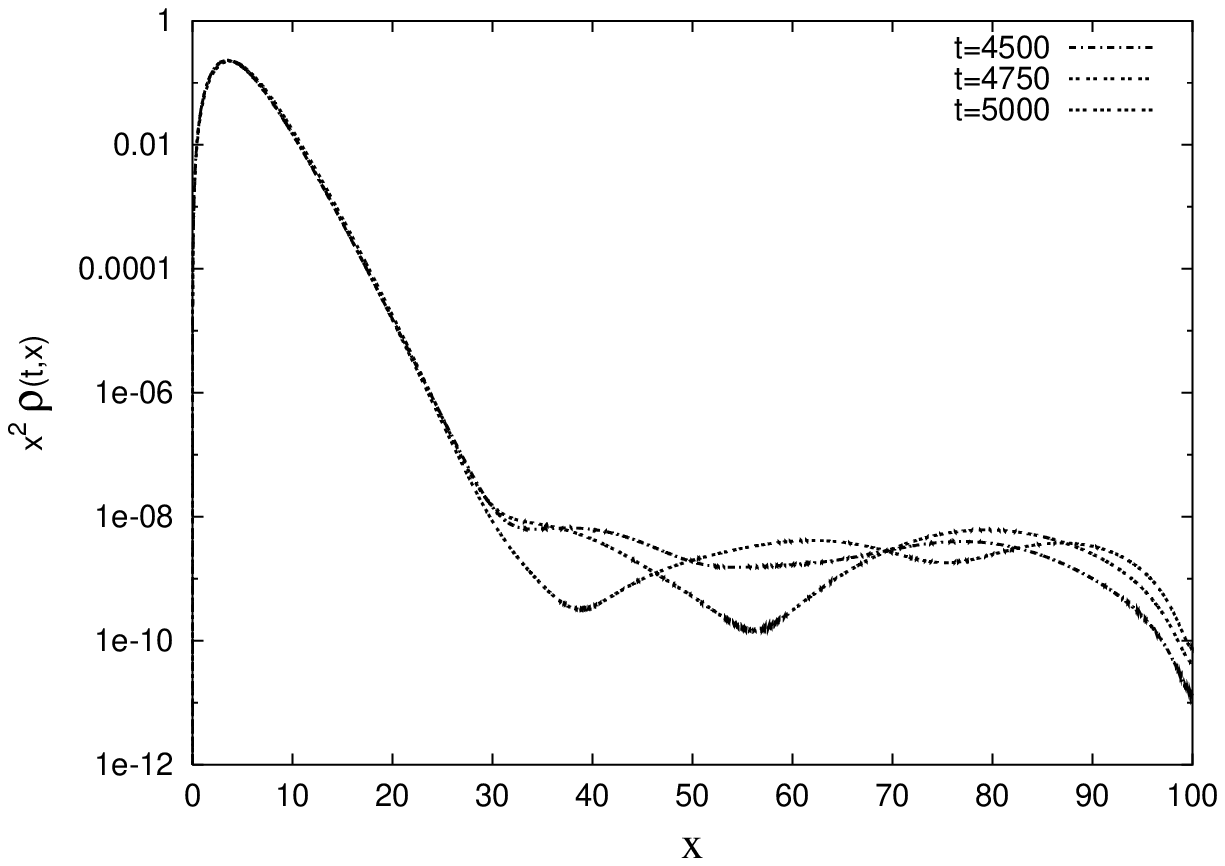}}
\caption{(Top) Maximum values of the radial metric function
$g_{rr}(t,x)=a^2(t,x)$. The initial configuration is that of
Fig.~\ref{fig:phi_s} with $\Delta x=0.01, \Delta t/ \Delta x =
0.5$. The boundary is at $x=100$ and the evolution is shown to
$t=5,000 \, m^{-1}$, some $50$ crossing times. In fact, we have
followed the run up to $t=20,000 \, m^{-1}$ and no change was
observed. Thus, we can conclude that the initial configuration is
stable against {\it small} perturbations.  (Bottom) The density
profile $\rho_\Phi(t,x)$ times $x^2$ is plotted for late times.}
\label{fig:a_ec}
\end{figure}

\begin{figure}[htp]
\centerline{\epsfysize=6cm \epsfbox{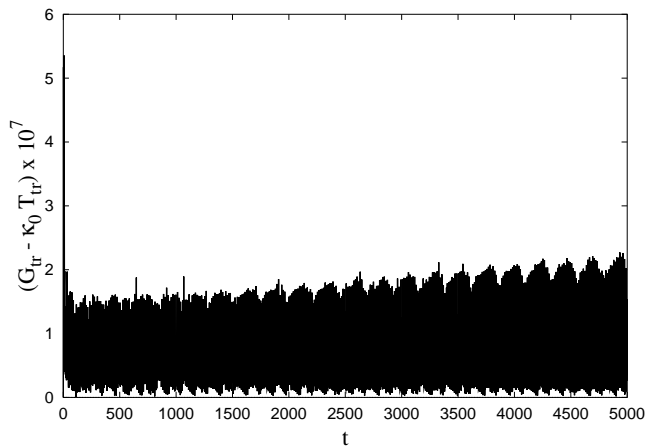}}
\caption{$L2$ norm of the momentum constraint for the numerical
evolution of the oscillaton shown in
Figs.~\ref{fig:phi_s},\ref{fig:a_ec}}.
\label{fig:adot_ec}
\end{figure}

\begin{figure}[htp]
\centerline{ \epsfysize=6cm \epsfbox{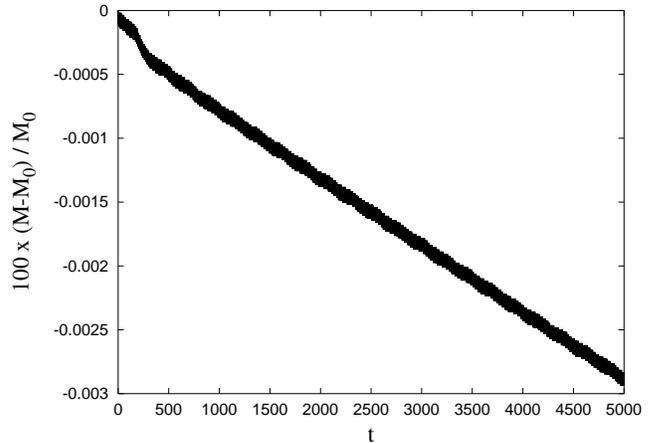}}
\caption{Total integrated mass as a function of time during the
evolution of the same oscillaton. We see that the oscillaton ejects a
very small amount of matter at around $t=300m^{-1}$. The overall
linear decrease is due to the numerical dissipation still present in
the code.}
\label{fig:mass}
\end{figure}

As a typical example of a S-branch oscillaton, we show in
Fig.~\ref{fig:a_ec} the evolution of the initial profile shown in
Fig.~\ref{fig:phi_s}, characterized by the central value
$\phi_1(0)=0.2828$. From the figure we see that such an oscillaton
is stable, maintaining the same oscillatory pattern from $t \simeq
300$ (see Fig.~\ref{fig:adot_ec}) up to times $t \simeq 20,000$
(in the plot we are showing only a fraction of the run).  That the
evolution is stable can also be seen from Fig.~\ref{fig:adot_ec},
in which we show the accuracy of the numerical evolution by
plotting the momentum constraint. In Fig~\ref{fig:mass} we show
the evolution of the total integrated mass.  We can see a small
appreciable adjustment of the original mass at around $t=300$,
indicating a small ejection of scalar field from the system. The
overall linear decay of the mass can be shown to be consistent
(using convergence tests) with a small amount of numerical
dissipation still present in our numerical method, and is
therefore not an intrinsic decay of the oscillaton.

The simulations show that the evolved profile oscillates periodically
at two different time scales. In particular, there is a short-period
oscillation, which we may identify with the fundamental oscillation,
and an overall vibration with a longer
period~\cite{seidel91a}. Following the literature of boson
stars~\cite{seidel90a}, we shall refer to the later as the
``quasi-normal modes''.

\begin{figure}
\centerline{\epsfysize=6cm \epsfbox{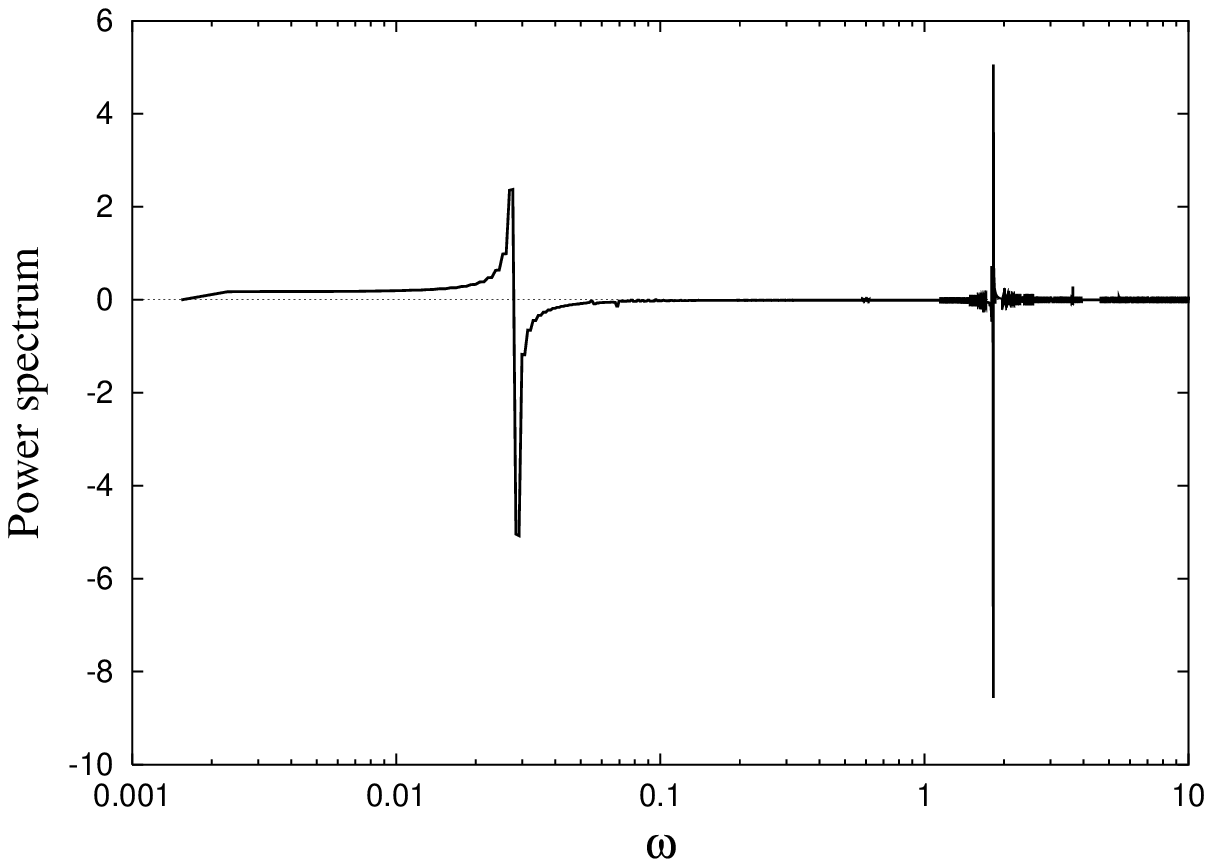}}
\centerline{\epsfysize=6cm \epsfbox{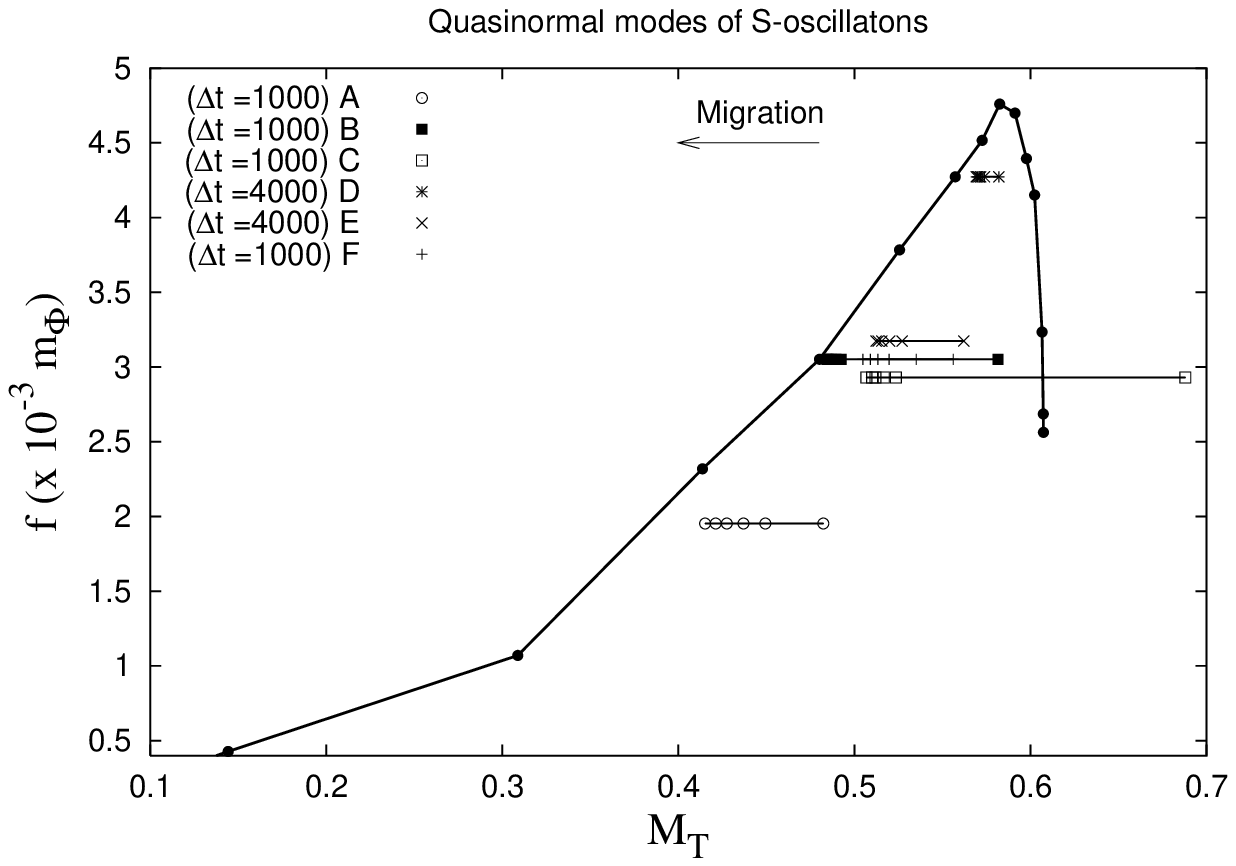}}
\caption{(Top) The power
spectrum of the evolution of the metric coefficient $g_{rr}$ for
the case in Fig.~\ref{fig:a_ec}. The configuration vibrates with
the quasi-normal frequency $(f/m)=4.52 \times 10^{-2}$ while
oscillates with even-multiples of the fundamental angular
frequency $(\omega/m) = 0.9119$.  The latter coincides with the
result of the eigenvalue problem, see Fig.~\ref{fig:phi_s}.
(Bottom) The frequencies of the quasi-normal modes obtained from
the evolution of slightly perturbed S-branch oscillatons. We also
show the migration of perturbed S-branch and U-branch oscillatons,
labelled {\tt A-F}; the details of their evolution can be found in
sections~\ref{four:i},\ref{five}. In the plot, $\Delta t$
represents the time intervals at which the evolved mass was
measured for each case.} \label{fig:freq}
\end{figure}

To determine the oscillatory scales of the solution, we calculated the
power spectrum of the evolution of $(g_{rr})_{max}$ in
Fig.~\ref{fig:a_ec}, and plotted it in Fig.~\ref{fig:freq}. Notice
that the dominant frequencies are the quasi-normal one and (twice) the
fundamental angular frequency, which are, respectively, $(f/m)=4.52
\times 10^{-2}$ and $(\omega / m) = 0.9119$. Higher modes
corresponding to even-multiples of the fundamental frequency are also
present, but their power is smaller. Observe that the fundamental
frequency is a little bit readjusted too, but its value is consistent
with the eigen-solution in Fig.~\ref{fig:phi_s}.

We systematically evolve all other S-branch equilibrium configurations
shown in Figs.~\ref{fig:mass_s}, calculating their corresponding power
spectrum and then their quasi-normal and fundamental frequencies. In
all cases, the evolved profiles behaved accordingly with the
description given above for the (representative) case
$\phi_1(0)=0.2828$.

For instance, the values of the fundamental frequencies were just a
little bit readjusted in all cases, but its values were always
consistent with those shown in Figs.~\ref{fig:mass_s}. In general, the
power spectrum of more dilute oscillatons is dominated by the
fundamental frequency, while the quasi-normal frequency dominates for
oscillatons near the critical point. Also, the linear numerical
dissipation of mass was as in Fig.~\ref{fig:adot_ec}, suggesting also
that it should be attributed to the numerical method and not to the
intrinsic properties of oscillatons.

The resulting plot $f \, vs \, M$ in Fig.~\ref{fig:freq} has been very
useful to analyze the evolution of boson
stars\cite{seidel90a,seidel98a}, and we will see it is useful for
oscillatons as well. The plot shows a maximum and a sharp decline near
the critical mass. This is a typical behavior indicating the transition
from stable to unstable configurations.

\begin{figure}[htp]
\centerline{ \epsfysize=6cm \epsfbox{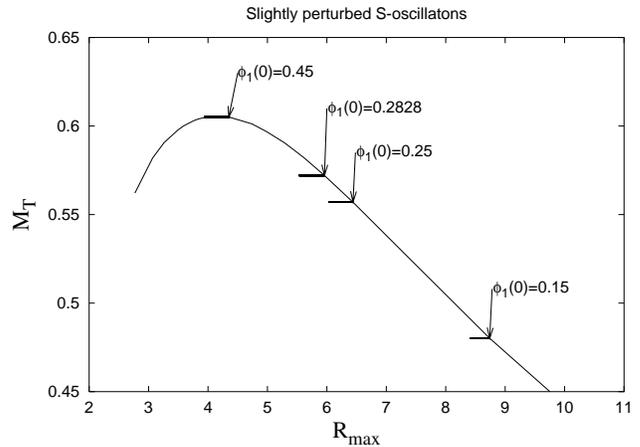}}
\caption{The evolution of the total mass $M_T$ and $R_{\rm max}$ is
shown for different S-oscillatons. The runs were followed up to
$t=5,000 \, m^{-1}$. It can be clearly seen that the evolutions
proceed just by small oscillations around the initial
profiles. Actually, since the profile in Fig.~\ref{fig:mass_s} uses
the initial value $R_{max}(0)$, the oscillations are shifted to the
{\it left} of the solid line. Hence, it is manifest that these
oscillatons are not migrating nor decaying. For a comparison with
truly migrating oscillatons, see
Figs.~\ref{fig:mvsr_sp},\ref{fig:mvsr},\ref{fig:mvsr_up}.}
\label{fig:mvsr_ec}
\end{figure}

`To further demonstrate that the S-oscillatons are stable and that the
frequencies shown in Fig.~\ref{fig:freq} are their intrinsic
quasi-normal modes, we show in Fig.~\ref{fig:mvsr_ec} the evolution of
the total mass $M_T$ and $R_{max}$ compared to the equilibrium
configurations as shown in Fig.~\ref{fig:mass_s}. It is clear that the
slightly perturbed S-oscillatons oscillate with very small amplitudes
around the original equilibrium configurations. In other words, they
are not migrating to other oscillatons nor decaying. This can be
compared to the migrating oscillatons shown in
Figs.~\ref{fig:mvsr_sp},\ref{fig:mvsr},\ref{fig:mvsr_up} below.

The results presented above give evidence for the points (i-ii)
outlined at the beginning of this section. We would like to stress
here that the important result is that there do exist stable
oscillatons (at least in the S-branch). As a side-effect, we have also
proven the consistency between the eigenvalue problem and the
numerical evolution code.


\subsubsection{Perturbed S-oscillatons}
\label{four:i}

Our interest now is to determine whether S-branch oscillatons are
stable against {\it strong} perturbations. That is, we want to know
the conditions under which such oscillatons will either collapse into
a black hole, disperse away or form another oscillaton. The latter is
a rather interesting possibility since, as we shall show below, it
implies the migration of oscillatons.

\begin{figure}[htp]
\centerline{ \epsfysize=6cm \epsfbox{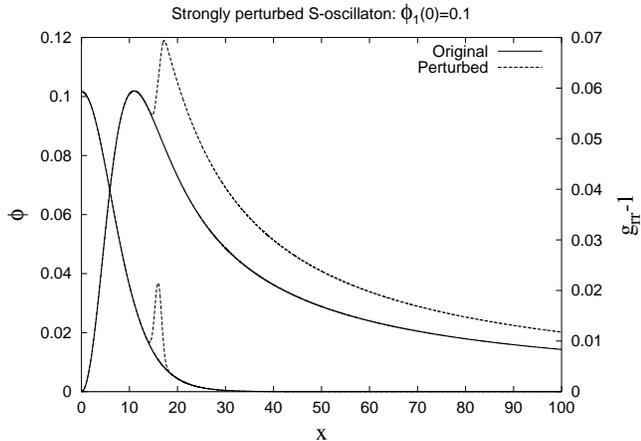}}
\caption{Examples of the scalar field and $g_{rr}$ profiles of a
strongly perturbed S-oscillaton by means of a Gaussian
perturbation. Shown is the case of a $\phi_1(0)=0.1$-oscillaton whose
original mass was increased $40$\%. The evolution of this oscillaton
is given in Figs.~\ref{fig:freq} (label {\tt
B}),\ref{fig:1m15},\ref{fig:mvsr_sp}. See also text below for
details.}
\label{fig:gauss}
\end{figure}

\begin{figure}
\epsfysize=6cm
\epsfbox{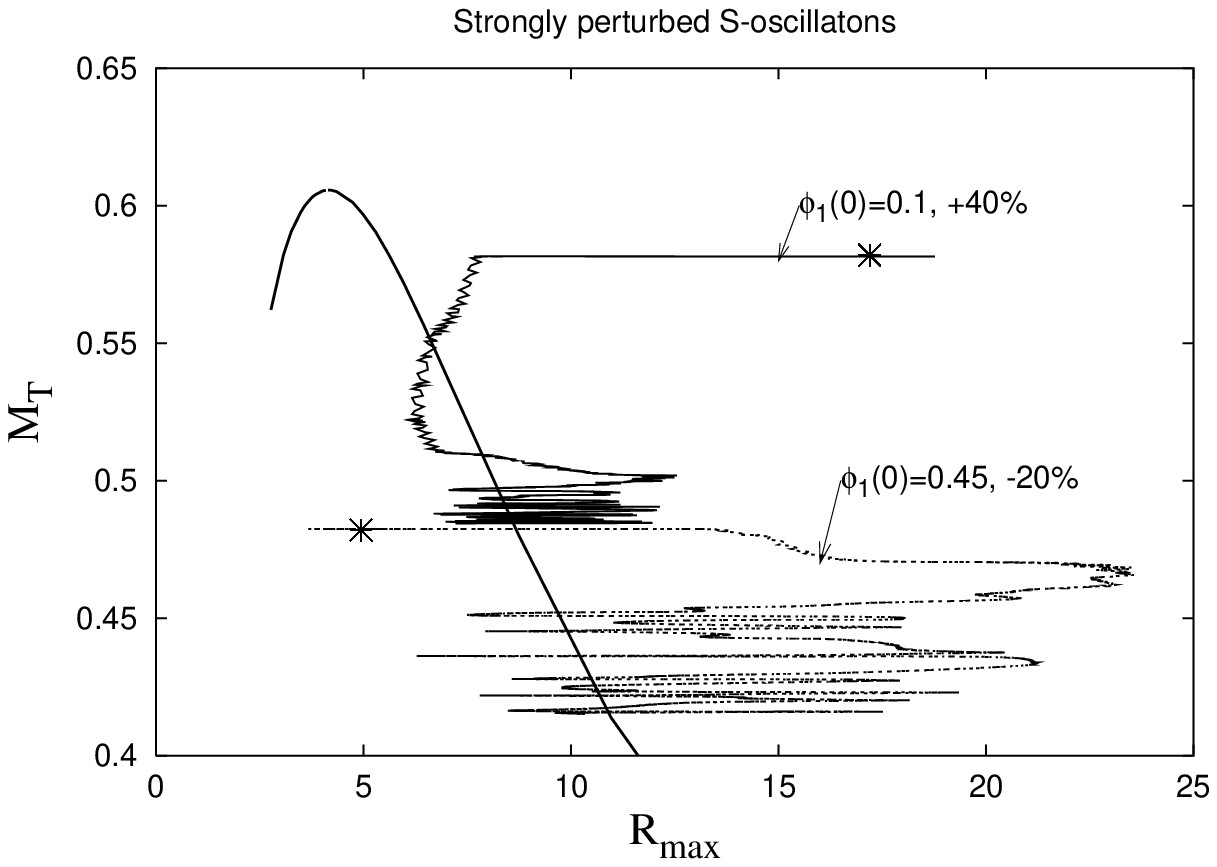}
\epsfysize=6cm
\epsfbox{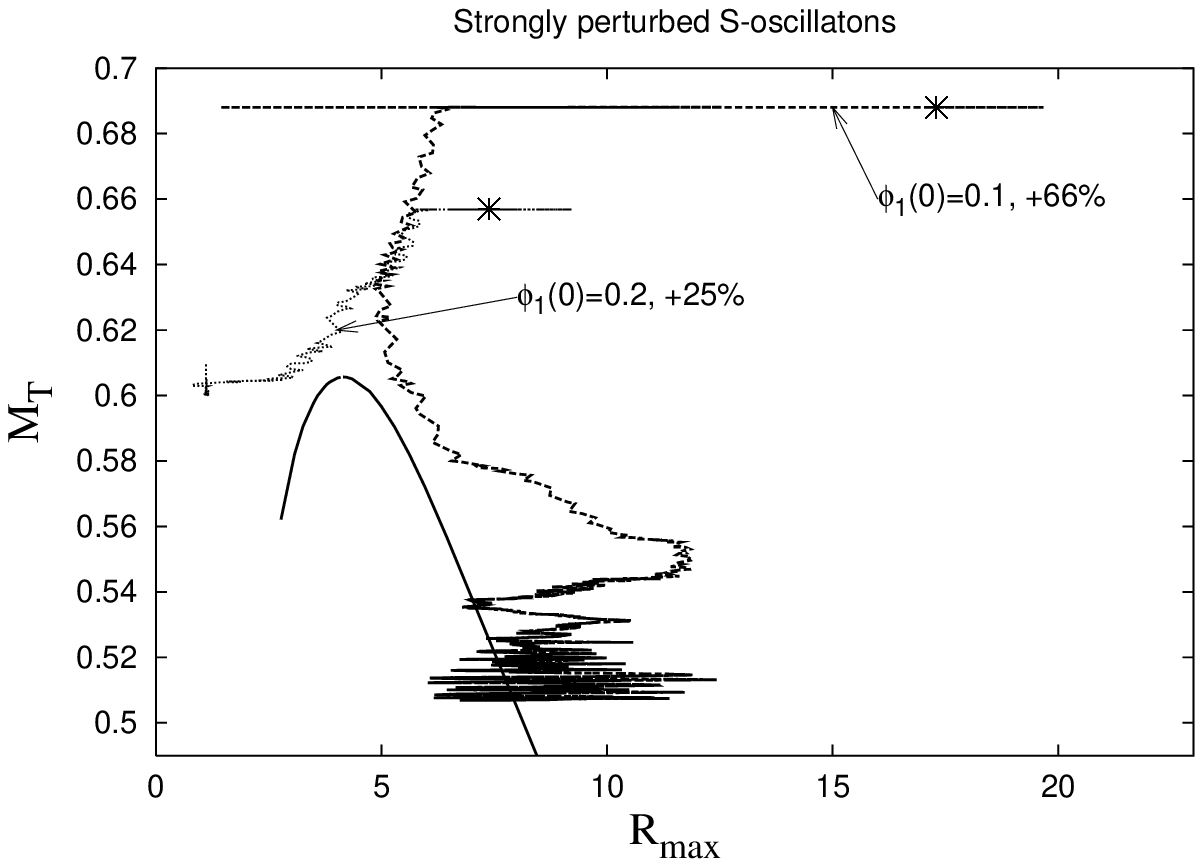} \caption{Evolutions of
different strongly perturbed S-oscillatons, see text for details
and cases labelled {\tt A-C} in Fig.~\ref{fig:freq}. The
percentage figures are the perturbations made in their original
masses. In general, a perturbed S-oscillaton migrates back to the
S-branch if its initial mass is less than the critical value
$M_c=0.606$. The migration proceeds by the loss of mass, through
the so-called gravitational cooling mechanism. This mechanism is
so efficient in the case of dilute oscillatons, that it can
prevent the formation of a black hole even if the initial mass is
larger than the critical value. However, this cooling mechanism is
sometimes not enough for dense oscillatons. The asterisks denote
the starting points.  The runs were followed up to $t=5,000
m^{-1}$.} \label{fig:mvsr_sp}
\end{figure}

We have included Gaussian-like perturbations in the original
equilibrium configurations, like that shown in Fig.~\ref{fig:gauss},
which can be seen just as perturbations to the original mass of the
oscillatons. (We shall call {\it original} the profiles shown in
Figs.~\ref{fig:mass_s} and their corresponding parameters). The main
results are summarized as follows. iii) If an oscillaton is perturbed
in such a way that its mass is less than the critical mass $M_c \simeq
0.606 \, (m^2_{\rm Pl}/m)$,\footnote{This value is just $0.16$\% lower
than the one determined from the eigenvalue problem in
section~\ref{two:i}. See also section~\ref{five:ii}.} it migrates to
another solution in the S-branch. iv) On the other hand, if the
initial mass is larger than $M_c$, the oscillaton can either migrate
to an S-oscillaton or collapse to a black hole. The collapse to a
black hole can be prevented by the oscillaton ejecting the mass
excess, the so-called gravitational cooling
mechanism~\cite{seidel94a}. This mechanism is highly effective for
dilute oscillatons. In both cases (iii,iv) above, the evolution of the
perturbed profiles can be tracked according to their vibration
frequencies, i.e., the quasi-normal modes. We have plotted some
typical evolutions of perturbed S-branch oscillatons in
Fig.~\ref{fig:mvsr_sp} which show the different behaviors described
above.

We start by discussing point (iii). In the first case, we decreased
the initial mass of a $\phi_1(0)=0.45$-oscillaton to an initial
value $M_i=0.484$ ($20$\% less than its original mass $M=0.605$).
We can observe that initially the oscillaton expands and then
bounces back while it looses mass, to finally settle down on the
S-branch. The migration is manifest in the trajectory shown in
Fig.~\ref{fig:mvsr_sp}, that oscillates around the stable
equilibrium configurations discussed in the previous section.
During the evolution, the oscillaton maintains a fixed vibration
frequency.  Its path of migration is labelled {\tt A} in
Fig.~\ref{fig:freq}, and shows that the oscillaton will stop at a
diluted one.

\begin{figure}
\epsfysize=6cm
\epsfbox{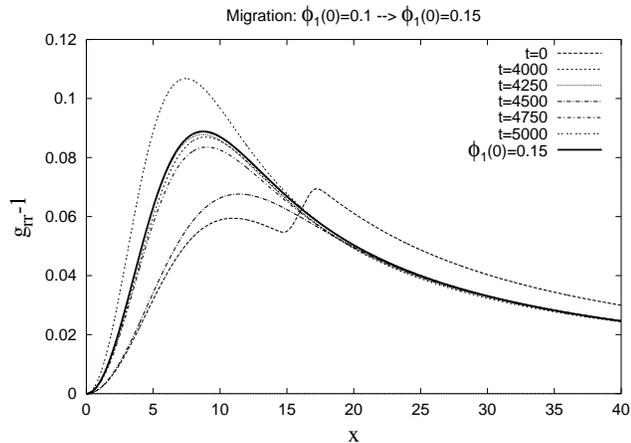}
\caption{Evolved profiles of the metric coefficient $g_{rr}$ for a
strongly perturbed $\phi_1(0)=0.1$-oscillaton (see also
Figs.~\ref{fig:gauss},\ref{fig:mvsr_sp}), which is migrating to a
$\phi_1(0)=0.15$-oscillaton. This is consistent with the path (labeled
{\tt B}) in Fig.~\ref{fig:freq}. The times and distances are given in
units of $m^{-1}$.}
\label{fig:1m15}
\end{figure}

A second case corresponds to a $\phi_1(0)=0.1$-oscillaton which
was strongly perturbed as to have an initial mass of $M_i=0.579$
($40$\% over its original mass $M=0.414$). Despite the strong
perturbation, the oscillaton collapses and looses enough mass to
settle down onto another S-oscillaton. By measuring the vibration
frequency, we find that the migrating oscillaton follows the path
labelled {\tt B} in Fig.~\ref{fig:freq}, which suggests that the
oscillaton is migrating to a $\phi_1(0)=0.15$-like profile. This
can also be seen in Fig.~\ref{fig:1m15}, in which the profile of
the metric coefficient $g_{rr}$ rapidly approaches and oscillates
around the final configuration.

To illustrate point (iv), a $\phi_1(0)=0.1$-oscillaton was perturbed
to have an initial mass of $M_i=0.687$, a value larger than the
critical one. However, we can observe that the oscillaton rapidly
loses mass and migrates back to the S-branch. Its migration path is
labeled {\tt C} in Fig.~\ref{fig:freq}. This is a typical example of
the efficiency of the gravitational cooling mechanism in enough dilute
oscillatons.

The opposite case is shown by a perturbed $\phi_1(0)=0.2$-oscillaton,
whose initial mass is $M_i=0.657$. The gravitational cooling mechanism
is not efficient enough in this case to prevent the formation of a
black hole. We see this by noticing that its trajectory stops at the
value of $R_{max}$ corresponding to the Schwarzschild radius of the
hole.  However, we have observed that the same oscillaton does migrate
to the S-branch if the initial mass is somewhat less perturbed, while
still being larger than the critical one. This is also a generic
phenomenon we have observed by perturbing other S-oscillatons. This
tells us that the gravitational cooling mechanism is highly efficient
if the scalar field is diluted enough, like in the cases shown in the
original paper of Seidel \& Suen~\cite{seidel94a}.

The results presented so far point out that, apart from being stable
configurations, S-branch oscillatons are indeed the final states in
the evolution of other perturbed S-oscillatons. These two properties
(stability and final state-quality) are the imprint of S-oscillatons.
To end this section, we would like to stress here the important role
of the quasi-normal modes to follow the migration process of
oscillatons. The examples presented so far show that the vibration
frequency of the system remains the same during its evolution, even
while the system looses mass.


\section{Evolution of oscillatons: The U-branch}
\label{five}

We shall call U-branch oscillatons those equilibrium configurations on
the right (left) hand side of the critical configuration in the plot
$M \, {\rm vs} \, \phi_1(0)$ ($M \, {\rm vs} \, R_{max}$) in
Fig.~\ref{fig:mass_s}. We also evolved these equilibrium
configurations, under the same idea that the eigen-solutions of
section~\ref{two:i} are already slightly perturbed configurations. The
main result of this section is that U-oscillatons are intrinsically
unstable since, as we shall see below, they decay and migrate to the
S-branch under {\it small} perturbations.

\begin{figure}
\epsfysize=6cm
\epsfbox{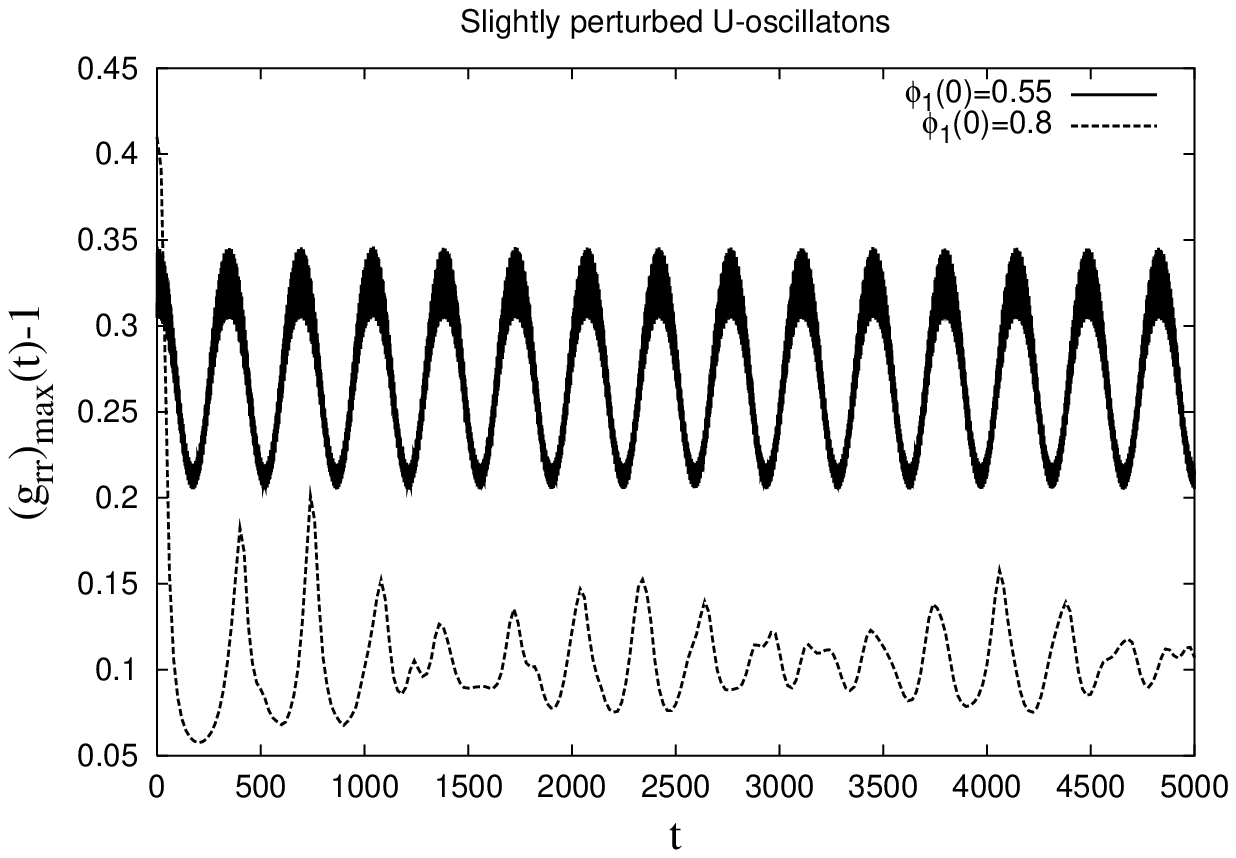}
\epsfysize=6cm
\epsfbox{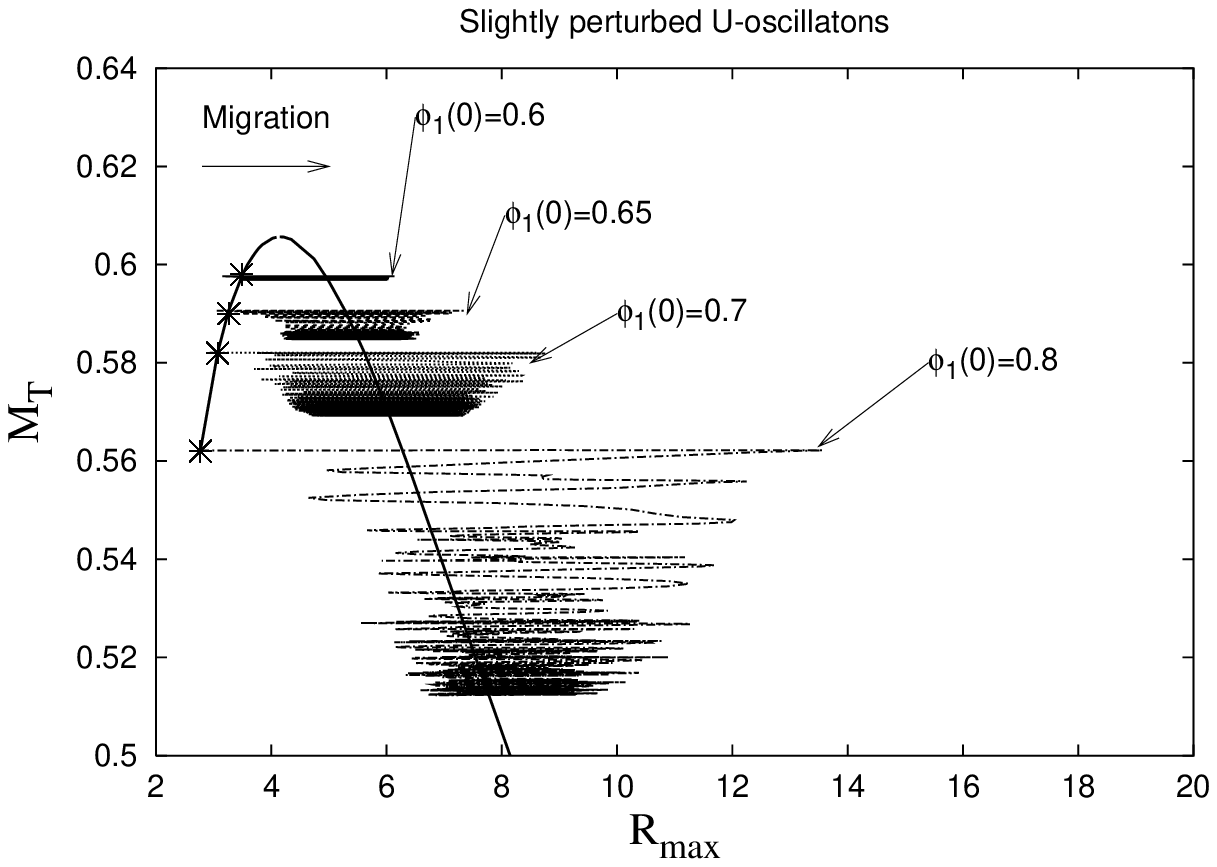}
\caption{(Top) Evolution of the maximum value of the radial metric
coefficient $g_{rr}$ for two slightly perturbed oscillatons in the
U-branch, whose corresponding central initial values are
$\phi_1(0)=0.55,0.8$.  Larger values of the latter mean more
instability. (Bottom) The migration of different U-oscillatons is
manifest in a graph of $M \, {\rm vs} \, R_{max}(t)$. By looking at
the plot $f \, {\rm vs} \, M$ in Fig.~\ref{fig:freq}, we can see at
which point the different oscillatons will stop at. For instance, the
labels {\tt D,E} correspond to the $\phi_1(0)=0.7,0.8$-oscillatons
shown in this graph, respectively. The time is in units of $m^{-1}$
and the asterisks denote the starting points.}
\label{fig:mvsr}
\end{figure}

Shown in Fig.~\ref{fig:mvsr} are some instances of slightly perturbed
U-oscillatons, and the plot $M \, {\rm vs} \, R_{max}$ speaks by
itself when compared to Fig.~\ref{fig:mvsr_ec}. The U-branch
oscillatons are intrinsically unstable; they migrate to and settle
down onto the S-branch, even under small perturbations. The larger the
central value ($\phi_1(0) > 0.48$), the more unstable they are. Their
migration to the S-branch also confirms the stability of the
S-oscillatons and their important role as the final states in the
evolution of migrating oscillatons.

As we have done before, the migration of oscillatons can be tracked by
determining their quasi-normal frequencies. For example, the slightly
perturbed $\phi_1(0)=0.7,0.8$-oscillatons shown in Fig.~\ref{fig:mvsr}
are labeled {\tt D,E}, respectively, in Fig.~\ref{fig:freq}.


\subsection{Perturbed U-oscillatons}
\label{five:i}

We have also perturbed the U-branch equilibrium configurations and
studied whether they migrate or collapse into black holes. The main
results of this section are as follows. i) If the initial mass
somewhat larger than the original mass of the eigen-configuration, the
oscillaton collapses into a black hole. This result is independent of
the kind of perturbation made and then it can be characterized by the
excess in mass only. ii) If the mass of the original configuration is
decreased by the perturbation, the oscillaton migrates to the
S-branch.

\begin{figure}
\epsfysize=6cm
\epsfbox{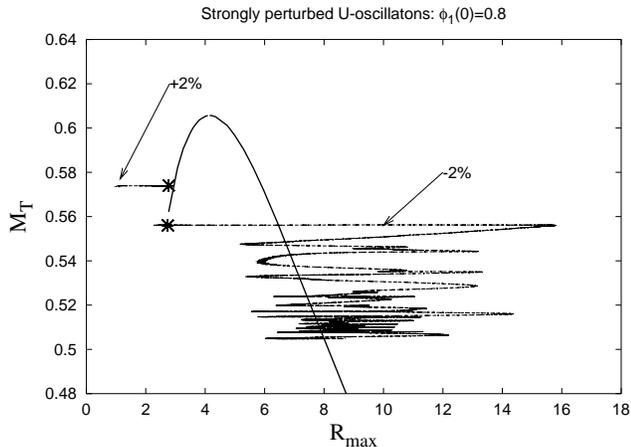}
\caption{The paths of the evolution of a perturbed
$\phi_1(0)=0.8$-oscillaton in a plot $M \, {\rm vs} \, R_{max}$. The
collapse into the black hole is represented by the horizontal line
going to the left, ending at the Schwarzschild radius of the final
black hole; see also Fig.~\ref{fig:BH}. The same oscillaton was also
perturbed by decreasing its original mass, in which case it migrates
and settles down into the S-branch, see text and Fig.~\ref{fig:8m15}
for details. The asterisks denote the starting points. The runs were
followed up to $t=5,000 m^{-1}$.}
\label{fig:mvsr_up}
\end{figure}

\begin{figure}
\epsfysize=6cm
\epsfbox{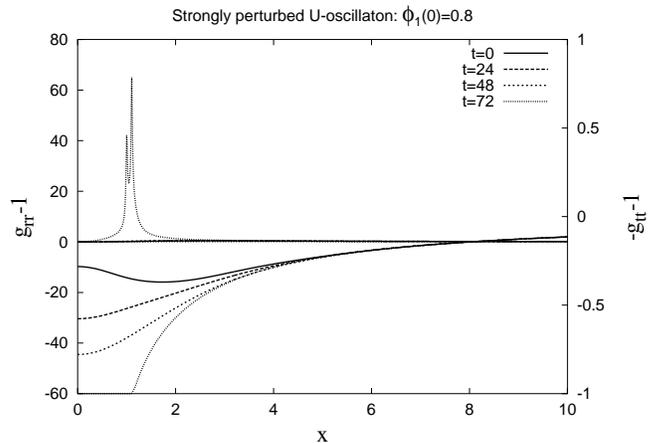}
\caption{Plots of the metric coefficients $g_{tt},g_{rr}$ for a
perturbed configuration with an excess mass of $2$\%. Even though the
value of the initial mass is well below $M_c \simeq 0.606$, the
oscillaton collapses into a black hole.}
\label{fig:BH}
\end{figure}

\begin{figure}
\epsfysize=6cm
\epsfbox{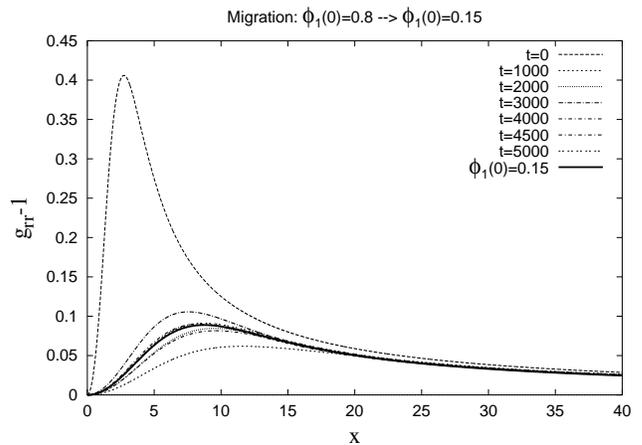}
\caption{Evolved profile of the metric coefficient $g_{rr}$ for a
$\phi_1(0)=0.8$-oscillaton migrating to a
$\phi_1(0)=0.15$-oscillaton. This is consistent with the graphs in
Fig.~\ref{fig:freq} (label {\tt E}) and in Fig.~\ref{fig:mvsr_up}. The
times and distances are given in units of $m^{-1}$.}
\label{fig:8m15}
\end{figure}

The $\phi_1(0)=0.8$-oscillaton is a typical example of U-oscillaton
and its evolution under strong perturbations is plotted in
Fig.~\ref{fig:mvsr_up}. If the initial mass of this oscillaton is
perturbed so that is $2$\% more massive that the original mass, the
final stage of the collapse is a black hole, even though the initial
mass is less than the critical value.  This can also be seen in
Fig.~\ref{fig:BH}, where we see plots of the metric coefficients
$g_{tt}$ and $g_{rr}$.  The coefficient $g_{tt}$ shows the well-know
``collapse of the lapse'' (the lapse drops to zero) characteristic of
black holes, while the coefficient $g_{rr}$ shows the ``grid
stretching'' effect (te radial metric grows exponentially), also
typical of black hole evolutions. On the other hand, if the initial
mass of the same oscillaton is $2$\% less than the original mass, the
oscillaton rapidly migrates to the S-branch and settles down into a
stable oscillaton. Its evolution path in Fig.~\ref{fig:freq} (label
{\tt F}), overlaps with that of the migrating S-oscillaton shown in
Fig.~\ref{fig:gauss},\ref{fig:1m15} (label {\tt B} in
Fig.~\ref{fig:freq}). Thus, the final state of the evolution seems to
be a $\phi_1(0)=0.15$-oscillaton, as can also be seen in
Fig.~\ref{fig:8m15}. This last fact reveals again the usefulness of
the plot $f \, {\rm vs} \, M$ to determine the final configuration of
a migrating oscillaton.

We would like to summarize here the results presented so far. We have
found that the imprint of U-branch oscillatons is that they are
intrinsically unstable.  That is, even small perturbations provoke
their migration to other configurations, instead of the small
oscillations around the original ones.  This latter fact again
confirms the stability and final state-status of S-oscillatons.
Another particular characteristic of U-oscillatons is that the
formation of a black hole is very likely if the original profile is
perturbed by adding mass to it. This phenomenon occurs even if the
perturbed initial mass is less than the critical one, so that the
gravitational cooling mechanism is not efficient at all in the
U-branch.


\subsection{The S-U transition point}
\label{five:ii}

Now we take a closer look at the equilibrium configurations near the
critical profile, the one being the most massive in
Figs.~\ref{fig:mass_s}, which we shall call the S-U transition point.

To begin with, let us recall the basic characteristics of the
S,U-branches.  The basic imprint of S-branch oscillatons is that they
are stable and then oscillate around the initial profiles when
slightly perturbed, see Fig.~\ref{fig:mvsr_ec}. On the other hand,
U-branch oscillatons are intrinsically unstable and they migrate to
the S-branch even when slightly perturbed, like those cases shown in
Fig.~\ref{fig:mvsr}.

\begin{figure}
\epsfysize=6cm
\epsfbox{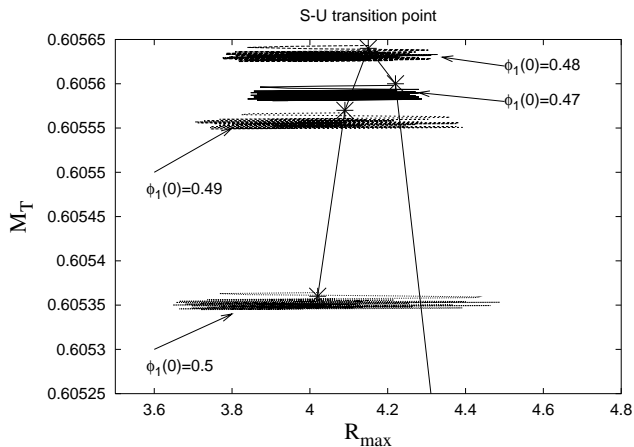}
\caption{Evolution of slightly perturbed configurations near the S-U
transition point, which is represented by the most massive equilibrium
configuration in Figs.~\ref{fig:mass_s}. The oscillations of
oscillatons to the right of the critical configuration are shifted to
the left of the solid line, as was the general case for the S-branch,
see Fig.~\ref{fig:mvsr_ec}.  On the other hand, the oscillatons to the
left of the critical configuration rapidly migrate to the S-branch and
oscillate like stable S-oscillatons, las in
Fig.~\ref{fig:mvsr}. Therefore, the critical configuration truly marks
the transition between the S and U- branches.}
\label{fig:mvsr_su}
\end{figure}

The evolution of the profiles near the S-U transition point when
slightly perturbed are shown in Fig.~\ref{fig:mvsr_su}. We observe
that the results summarized above are again confirmed, and then the
critical configuration characterizes the transition between the S and
U branches under {\it small} perturbations. We have also found the
same results in the case of strong perturbations. If the mass of these
oscillatons is decreased, they migrate to S-oscillatons. If the mass
is increased even by a small amount, they collapse rapidly forming
black holes.

That U-oscillatons form black holes is not surprising.  The fact
that S-oscillatons can also collapse to black holes near the
critical point is a consequence of the inefficiency of the
gravitational cooling mechanism in the case of dense
S-oscillatons, as it was found in section~\ref{four:i}. Actually,
if the mass of an S-oscillaton near the transition point is
perturbed to be larger than $M > 0.606$, it will collapse into
black holes; otherwise, it will migrate to more diluted
configurations. That is, the value $M_c \simeq 0.606$ can be seen
as the true critical value in the S-U transition region.

Thus, we can conclude that the critical oscillaton ($\phi_1(0) \simeq
0.48, \, M_c \simeq 0.606 m^2_{\rm Pl}/m$) marks the transition from the
S-branch to the U-branch.

\section{Conclusions}
\label{conclusion}

In this paper, we have studied the properties of oscillatons by numerically
evolving the Einstein-Klein-Gordon equations. According to the results,
oscillatons can be classified into two well definite groups, the S and
U-branches.

S-branch oscillatons are stable equilibrium configurations, which
oscillate according to their fundamental and intrinsic frequency
($\omega$). When slightly perturbed, these oscillatons vibrate with
definite frequencies which we have identified as the frequencies ($f$)
of their particular quasi-normal modes. This is supported by the fact
that these vibrations are small-amplitude oscillations around the
equilibrium configurations.  If these oscillatons are strongly
perturbed such that their original mass is decreased, they migrate to
and settle down onto other S-branch oscillatons.  This fact points out
that S-oscillatons should be seen as final states to which other
scalar configurations migrate to.

On the other hand, we have found that S-oscillatons do not, in
general, collapse into black holes if their mass is increased by the
perturbation.  Even though we have found that oscillatons with
high-central densities form black holes if their mass is larger than
the critical value $M_c \simeq 0.606$, diluted oscillatons can avoid
such fate by means of the gravitational cooling mechanism. The later
can be so efficient that strongly perturbed oscillatons can migrate to
stable configurations, even if their initial mass was much larger than
the critical value.

The evolutions of the U-branch oscillatons consistently show that they
are intrinsically unstable configurations. Even under small
perturbations, they migrate to the S-branch. This migration also
appears when they are strongly perturbed in such a way that their
original mass is decreased. However, another manifestation of their
unstable nature is that they rapidly collapse into black holes if
their mass is increased, even by small amounts and even if the initial
mass is well below the critical value.

In all cases of migrating oscillatons, we found that their evolution
can be followed by measuring their vibration frequency, and this can
be used to predict to which configuration the perturbed oscillatons
will migrate to.

Through this paper, we have studied oscillatons with a quadratic
scalar potential for simplicity. However, we have already
initiated similar studies to include a self-interaction in the
potential and they are work in progress to appear elsewhere.


\acknowledgments

We would like to thank Erasmo G\'omez and Aurelio Esp\'{\i}ritu for
technical support, and Edward Seidel for useful discussions. T.M. and
L.A.U. acknowledge the kind hospitality of the AEI, where part of this
work was developed. The simulations were performed in the Laboratorio
de Superc\'omputo Astrof\'{\i}sico del Departamento de F\'{\i}sica,
CINVESTAV.  This work was partly supported by CONACyT M\'exico, under
grants 010385 (L.A.U.), 32138-E and 34407-E. MA, and DN acknowledges
the DGAPA-UNAM grant IN-122002.




\end{document}